\documentclass[aps,prl,twocolumn,showpacs,superscriptaddress]{revtex4}
\usepackage{graphicx}
\usepackage{amsmath,amssymb,latexsym}
\bibstyle{apsrev.bib}

\begin{document}

\title{Topological signatures of globular polymers}
 
\author{M. Baiesi}
\affiliation{Dipartimento di Fisica, Universit\`a di Padova, Via Marzolo 8, I-35131 Padova, Italy}
\author{E. Orlandini}
\affiliation{Dipartimento di Fisica, Universit\`a di Padova, Via Marzolo 8, I-35131 Padova, Italy}
\affiliation{Sezione INFN, Universit\`a di Padova, Via Marzolo 8, I-35131 Padova, Italy}
\author{A. L. Stella}
\affiliation{Dipartimento di Fisica, Universit\`a di Padova, Via Marzolo 8, I-35131 Padova, Italy}
\affiliation{Sezione INFN, Universit\`a di Padova, Via Marzolo 8, I-35131 Padova, Italy}
\author{F. Zonta}
\affiliation{Dipartimento di Fisica, Universit\`a di Padova, Via Marzolo 8, I-35131 Padova, Italy}

\date{\today}

\begin{abstract}

Simulations in which a globular ring polymer with delocalized knots
is separated in two interacting loops by a slipping link, or in two 
non-interacting globuli by a wall with a hole, show how the minimal 
crossing number of the knots controls the equilibrium statistics. With 
slipping link the ring length is divided between the loops according to 
a simple law, but with unexpectedly large fluctuations. 
These are suppressed only for unknotted loops, whose length distribution shows 
always a fast power law decay. 
We also discover and explain a topological effect interfering with that 
of surface tension in the globule translocation through a membrane nanopore.

\end{abstract}


\maketitle

Most often, ring polymers are experimentally studied in 
situations in which their topological entanglement does 
not change in time. However, most of our theoretical 
and numerical understanding of polymer statistics relies on 
ensemble descriptions in which the rings assume all possible 
topologies~\cite{Orlandini&Whittington:2007:Rev-Mod_Phys}. 
An open challenge is that of determining up to 
what extent specific permanent entanglements in the form of 
knots or links~\cite{Adams:1994} can affect thermodynamic quantities and what 
is their possible role in determining 
peculiar behaviors when the polymer is subject to geometrical constraints 
interfering with the topology.

Different topologies are expected to determine 
different non-extensive corrections to the free energy
of a single ring in the limit of infinitely long chain. For 
example, the prime knot components of ring polymers in good 
solvent are weakly localized in this limit~\cite{Marcone:2005:J-Phys-A,Farago:2002:EPL}. As a 
consequence, if $N$ is the chain length, each component determines a correction
$\sim k_B \ln(N)/N$ to the entropy per monomer~\cite{Baiesi:2010:JSTAT,Baiesi:2010:PRE}. Indeed, each component 
behaves asymptotically as a point-like decoration which can place itself 
anywhere along the ring. Radically different conditions are realized 
below the theta temperature $T_{\theta}$~\cite{Vanderzande:1998}. Indeed, in the globular phase
knots are expected to delocalize. Numerical simulations indicate that the 
topological entanglement spreads on average on a portion of the ring 
whose length is proportional to $N$~\cite{Marcone:2007:PRE,Virnau:2005:J_Am_Chem_Soc}.
Fixed topology is also known to determine a finite size correction to the 
free energy of a globular ring, which is asymptotically negligible in 
comparison with that due to surface tension 
($\sim N^{-1/3}$). Indeed, at a temperature
$T \approx \frac 2 3 T_{\theta}$,  
the correction per monomer has been estimated~\cite{Baiesi:2007:PRL} 
as $ \sim C n_c^{a}N^{-2}$ where  $a\simeq 1.45$, 
$n_c$ is the minimal number of crossings of the 
knot~\cite{Adams:1994},
and $C$ is a remarkably large negative amplitude.  
Thus, $n_c$ qualifies 
as a topological invariant possibly relevant for the thermodynamics 
of a globular ring.

In the present work we face the challenge of elucidating 
the role of this invariant, by establishing some empirical
laws through which it rules the statistics of the globule
when suitable local geometrical constraints are imposed. These laws,
which generally hold in the presence of remarkably large fluctuations,
give a precise meaning to the notion of knot delocalization. 
We also discover novel thermodynamic phenomenona which are a
peculiar consequence of topology.

\begin{figure}
\includegraphics[width=0.9\columnwidth]{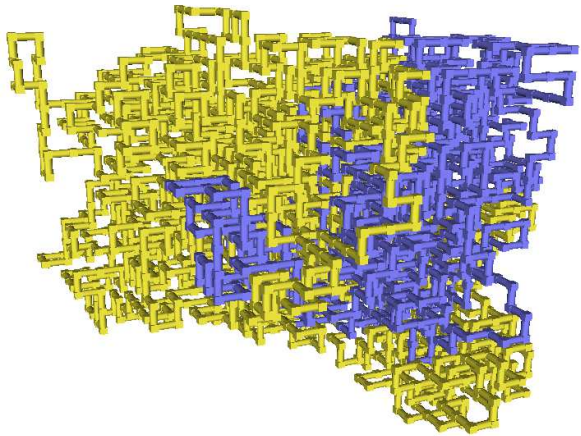}
\caption{$N=2040$ globule with slip link separating a $3_1$ knot in loop $1$ 
(yellow) from a $4_1$ knot in loop $2$ (blue).}
\label{fig:1}
\end{figure}

Suppose one forces a ring polymer to pass into a slipping link which 
divides it into two loops. The link is narrow enough to prevent the
passage of topological entanglement from one loop to the other. 
For instance, if the ring has a composite knot with two prime trefoil ($3_1$)
components~\cite{Adams:1994},
things can be arranged so that each fluctuating loop encloses one trefoil knot.
The loops are kept unlinked. We model flexible ring configurations 
as $N$-step self avoiding polygons on cubic lattice~\cite{Madras&Slade:1993}. An attractive 
interaction $J$ between nearest neighbor visited sites allows to obtain a 
collapsed globular phase for low enough temperature $T$~\cite{Vanderzande:1998} (Fig.~\ref{fig:1}). 
A Monte Carlo
simulation method adequate to preserve the ring topology is the grand-canonical 
BFACF one~\cite{Madras&Slade:1993}. However, staightforward application of BFACF to 
the globule meets a difficulty. Indeed, indicating by $K$ the step
fugacity and by $Z_N$ the canonical partition function, the grand canonical average 
$ \langle N \rangle = \Sigma_N N K^N Z_N / \Sigma_N K^N Z_N$ does not grow continuously
to $+ \infty$ upon approaching from below the critical value $K_c$ of
the fugacity. To the contrary, one gets evidence of a discontinuous
infinite jump right at $K=K_c$~\cite{Marcone:2007:PRE}.
We understand here this behavior in the light of the 
expected~\cite{Owczarek_et_al_PRL_1993,Baiesi:2007:PRL}
asymptotic form of $Z_N$:
\begin{eqnarray}
\ln(Z_N)=-F_N/kT \sim \text{const}+\mu N+\sigma N^{2/3}+\\
\nonumber
(\alpha -2) \ln(N)+\frac{C {n_c}^{a}}{N}
\label{eq:1}
\end{eqnarray}
where $\mu = -\ln(K_c)$, $\sigma <0$ is the interfacial tension, and
$\alpha$ is an unknown specific heat exponent that is expected to be
independent of topology. The last 
term on the r.h.s. of Eq.(\ref{eq:1}) is the above mentioned
topological correction to the total free energy $F_N$. 
The presence of the surface
term $\sim \sigma N^{2/3}$ implies that $ \langle N \rangle $ can not diverge 
continuously to $ + \infty$ for $K$ approaching $K_c$ from below.
In order to allow a continuous growth of $ \langle N \rangle $, we choose 
to multiply the usual grand-canonical weight of the ring configurations 
by a factor $\exp[(N-N_0)^2/v] $ which forces $N$ itself to fluctuate
around a value close to $N_0$ if $v$ is chosen small enough.
Using still BFACF moves the simulation augmented with the new statistical 
weight allows a quasi-canonical sampling of configurations with 
$N$ close to the value determined also by $N_0$ and $v$.
To improve sampling efficiency, we also implement a multiple Markov chain 
scheme~\cite{Tesi_et_al:1996:JSP} over different values of $N_0$ 
keeping $K$ fixed. Nevertheless, the simulations need a long CPU time (months) 
to sample a consistent statistics. For this reason we choose to sample only
at the temperature $T=2.5J $, as in~\cite{Baiesi:2007:PRL}.

\begin{figure}
\includegraphics[width=0.98\columnwidth]{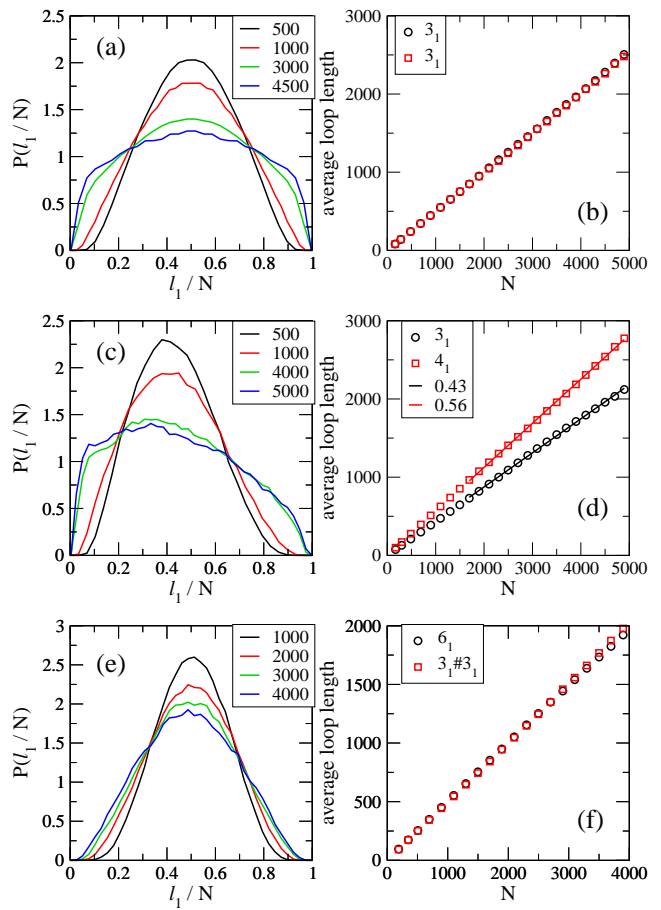}
\vskip 2mm
\caption{(a) Histograms of $P(l_1/N)$ for $3_1$ vs $3_1$. 
Different curves correspond to different $N$ values (see legend). 
(b) $ \langle l_1 \rangle _N$ (circles) and $ \langle l_2 \rangle _N$ (squares)  
as a function of $N$. 
(c)~Histograms of $P(l_1/N)$ for $3_1$ vs $4_1$ and (d)~corresponding
average loop lengths.
(e)~Histograms of $P(l_1/N)$ for $6_1$ (loop $1$) vs $3_1\#3_1$ and (f)~corresponding
average loop lengths.
}
\label{fig:2}
\end{figure}

Extensive simulations of the $3_1$ vs $3_1$ configuration allow 
to sample for various restricted ranges of 
$N$ the probability density function (PDF) of $l_1/N$, $P(l_1/N)$, where $l_1$ is the 
fluctuating length of one of the loops (Fig.~\ref{fig:2}a ). Remarkably, 
even for large $N$ the histogram does not seem to present the
bimodal shape found in the good solvent case and indicating a
dominance of configurations with large unbalance between $l_1$ and 
$l_2$~\cite{Marcone:2005:J-Phys-A,Marcone:2007:PRE}. 
To the contrary, an asymptotically flat histogram seems compatible with the 
data. Thus, while the two loops on average share equally the total ring 
length (Fig.2b), there are very broad fluctuations because all
possible partitions of the total ring length are almost equally 
probable.

\begin{figure}
\includegraphics[width=0.98\columnwidth]{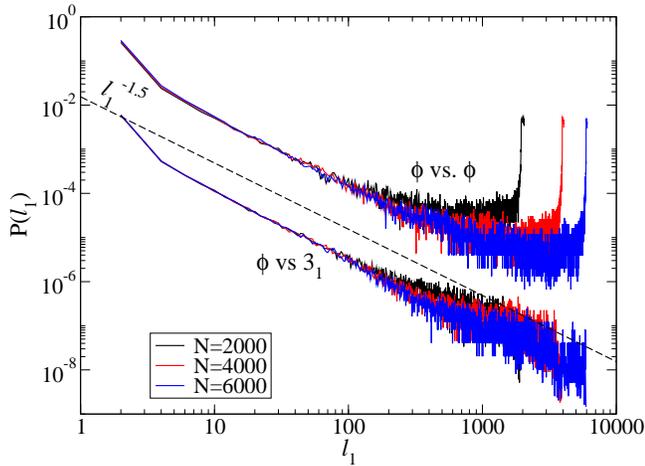} 
\caption{Log-log plots of $P(l_1)$ for $\emptyset$ vs $\emptyset$ 
(upper curves, for three values of $N$),
and $\emptyset$ vs $3_1$ (lower curves, shifted one decade down). 
The dashed line represents a power-law $l_1^{-1.5}$.
Each $N$ includes the statistics in the interval $[N-50,N+50]$: this does not 
alter the power-law tail and increases the statistics considerably.
}
\label{fig:3}
\end{figure}

To gain further insight, we simulate also
the case in which the loops are both unknotted. In this case strongly
unbalanced situations are clearly fovoured. The PDF of the length of the
smaller loop, say $l_1$, shows a power law decay $\sim l_1^{-x}$, with $x =1.55 \pm 0.04$
(Fig.~\ref{fig:3}). 
This suggests that the metric exponent of this loop is 
$\nu = x/d \approx 0.52 $. A value of $\nu$ close to $1/2$ is consistent with the
expectation of Gaussianity of a collapsed chain on relatively small length 
scales~\cite{Grosberg1994}.
When for example only loop $2$ has a $3_1$ knot, while loop $1$ is 
unknotted ($\emptyset$), we observe that $l_1$ never grows substantially 
compared to $N$, and the same power law behavior holds for the PDF 
of $l_1$ (Fig.~\ref{fig:3}). 
This power-law, implying that $\langle l_1 \rangle \sim N^{0.45(4)}$, denotes a weak localization
of the unknotted loop.

The above results are fully consistent with delocalization of the
$3_1$ knot inside the ring, since in no case a loop containing
the knot displays a stable regime in which its average length is a vanishing
fraction of $N$. Similar results hold if $3_1$ is replaced by another prime knot.
We further examine a competition $3_1$ vs $4_1$. The 
plots reported in Fig.~\ref{fig:2}c show that also in this case the
lengths of the loops keep fluctuating very broadly for increasing $N$, while $P(l_1/N)$
is not symmetric with respect to $l_1/N=1/2$ anymore. Quite remarkably, in 
this and similar competitions ($3_1$ vs $7_1$, $4_1$ vs $6_1$ etc.), 
a simple law is well obeyed by the canonical
averages $ \langle l_1 \rangle _N$ and $ \langle l_2 \rangle _N$:

\begin{equation}
 \langle l_i \rangle _N = N \frac{n_{ci}}{n_{c1}+n_{c2}},  \qquad i=1,2.
\label{eq:2}
\end{equation}

In this case the loop with the $3_1$ knot obtains on average
a fraction of the chain length equal to $3/7\approx 0.43$, while 
$4/7\approx 0.56$ go to the loop with $4_1$ knot. This is evidenced in
Fig.~\ref{fig:2}d by the linear fits of the average
loop lengths as a function of $N$.

The law in Eq.(2) highlights the key role played by the
topological invariant $n_c$ in the delocalized regime.
This role is further enphasized by considering the competition of
two knots that are different but with the same $n_c$. 
Fig.~\ref{fig:2}e and ~\ref{fig:2}f show the results for  $3_1 \# 3_1$  vs $6_1$. 
We see that, while the fluctuations remain very broad, 
also in this case the $ \langle l_i \rangle _N$'s still obey Eq. (\ref{eq:2}).
In addition the shape of the histograms is quite symmetric as in the $3_1$ vs $3_1$
case. However, the possible convergence towards a flat histogram is definitely slower
than in that case. These results suggest that the number of prime components of a 
knot is not a relevant invariant, unlike in the swollen regime.

\begin{figure}
\includegraphics[width=0.98\columnwidth]{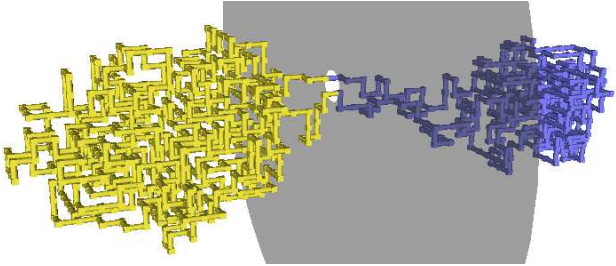}
\caption{Example of a wall--globule system.  
The hole (white) separates the $N=1000$ chain into two globuli with $3_1$ knot
(colored differently).}
\label{fig:4}
\end{figure}

\begin{figure}
\includegraphics[width=0.98\columnwidth]{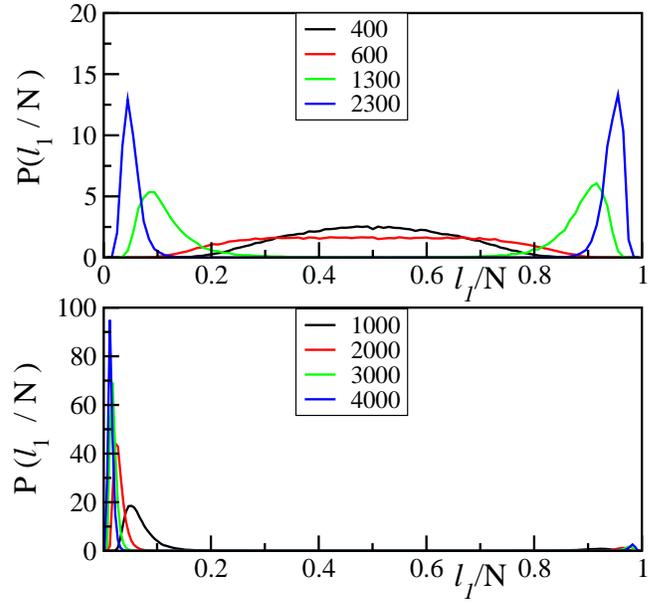}
\vskip 4mm
\caption{Histograms of $P(l_1/N)$ for $3_1 \#3_1$ (loop $1$) vs $6_1$
(top panel) and for $3_1$ (loop $1$) vs $4_1$ (bottom panel) when the two
globular loops are separated by an impenetrable wall. 
Different curves correspond to different $N$ values. 
}
\label{fig:5}
\end{figure}

In experiments the slipping link could be, e.g., a short
(unknotted) ring which is not linked to the knotted one, but 
just constrains it to pass through its interior. To discover further
consequences the topology may have  in the globular phase we replace the 
slipping link by a sufficiently narrow hole in an impenetrable wall,
Fig.~\ref{fig:4}. This schematizes the translocation of a ring polymer 
through a membrane or solid state nanopore, a phenomenon that, thanks to 
recent progress in nanotechnology  can be investigated 
experimentally~\cite{Kasianoviz:1996:PNAS,Smeets:2006:NanoLett}.
A related biological issue is the
effect that knots may have on the ejection process of packed viral DNA into the host
cell~\cite{Marenduzzo:2009:Proc-Natl-Acad-Sci-U-S-A:20018693,Matthews:2009:Phys-Rev-Lett:19257792}. 
Due to the presence of the wall the two loops do not  interact anymore and
can be considered as two independent, knotted globuli in competition. 
As shown in Fig.~\ref{fig:5} 
for a case with two different knots having however the same $n_{c1}=n_{c2}=6$,
$P(l_1/N)$ develops two remarkably symmetric peaks for $N\gtrsim 1000$, 
which become separated by a very high free energy 
barrier already at $N \approx 2300$, the maximum $N$ for which our
multiple chain sampling works efficiently. Similar results are obtained
for the simpler case $3_1$ vs $3_1$.

The situation turns out to be totally different when $n_{c1} \neq n_{c2}$. In 
these cases we observe an asymmetry, indicating a clear dominance of 
the globular loop hosting the knot with higher $n_c$. 
This dominance becomes more and more pronounced as $N$ increases.
We report the results for the case $3_1$ vs $4_1$
with rings of length up to $N = 4000$ in Fig.\ref{fig:5}.

How topology can produce the asymmetry manifested by
the above results for the wall case can be explained  as follows. 
The equilibrium share of the ring length between 
the two globuli should stably minimize the free energy. 
Due to independence, the total free energy is simply the 
sum of the free energies of the two globuli. 
We can assume that in the relevant ranges explored by our simulations
both $l_1$ and $l_2$ are sufficiently large for Eq.(\ref{eq:1}) to give a
reasonable approximation of the globule free energy. The surface term 
in $F_N$, if taken into account as the only correction, 
implies that both configurations in which one of the globuli has 
length $\lesssim N$, while the other is close to minimal, realize
equivalent stable minima of the total free energy with constraint
$l_1 +l_2 =N$. Let us indicate by $m_1$ and $m_2$
the value of $l_1$ and $l_2$, respectively, in the stable
configuration in which it happens to be the shorter loop.
From Eq.(\ref{eq:1}) one gets for the ratio of canonical probabilities
of these configurations:
\begin{eqnarray}
\nonumber
\ln\left(\frac{P(l_1=N -m_2, l_2=m_2)}{P(l_1= m_1, l_2=N-m_1)}\right)=\\
\nonumber
\sigma \left[(N-m_2)^{2/3} +m_2^{2/3} -(N-m_1)^{2/3}- m_1^{2/3}\right]+\\
\nonumber
(\alpha -2)\ln\left(\frac{(N-m_2)m_2}{(N-m_1)m_1}\right)+\\
C\left[\frac{n_{c1}^{a}}{N-m_2} +\frac{n_{c2}^{a}}{m_2} - \frac{n_{c1}^{a}}{m_1}
-\frac{n_{c2}^{a}}{N-m_1}\right]
\end{eqnarray}
In the case $n_{c1}=n_{c2}$ the presence of the topological correction 
does not matter, and thus we expect to see with the same
frequency a dominance of either one of the two globuli. 
The only topological effect left could be due to a slight difference
between $m_1$ and $m_2$ in the cases in which the two knots
in $1$ and $2$ are not identical. However, this difference
should be extremely small. Indeed, for example
the minimal lengths of lattice knots with the same $n_c$,
which are lower bounds for $m_1$ and $m_2$, 
are known to be almost coincident~\cite{Rensburg&Promislow:1995:JKT}. 
When we instead have 
$n_{c1} \neq n_{c2}$, the two configurations favored by surface 
tension are not equally probable anymore. The term apt to 
determine a substantial deviation of the quantity in Eq.(3) from 
zero originates in fact from the topological correction:

\begin{equation} 
C\left[\frac{n_{c2}^{a}}{m_2}- \frac{n_{c1}^{a}}{m_1}\right]
\label{eq:4}
\end{equation}
The main reason for this deviation is the remarkably large 
absolute value of $C$, which amplifies any even 
slight difference
$\frac{n_{c1}^{a}}{m_2}-\frac{n_{c2}^{a}}{m_1}$.
To get an estimate of the quantity in Eq.(\ref{eq:4})
for the first knots, we calculated it
for $C \approx -170$ and $a \simeq 1.45$, the values found for $T/J=2.5$~\cite{Baiesi:2007:PRL}, and $m_1$ and $m_2$
replaced by the existing estimates of the minimal lattice 
knot lengths~\cite{Rensburg&Promislow:1995:JKT}, consistently with the peak structures reported 
in Fig.\ref{fig:5}.
For the case $3_1$ vs $4_1$, the value of (\ref{eq:4})  turns out to 
be $\approx 7.5$.

Thus, a topological mechanism explains why 
a stable configuration is that in which the globule
with smaller $n_c$ is reduced to a minimal size,
while the other takes most of the ring length. This novel 
phenomenon should be relevant, e.g., for a slow dynamics of
translocation of globular knotted ring polymers through
membrane pores.

Putting things in perspective, althought
necessarily limited to the less complex knots, our results
are accurate enough to clearly qualify $n_c$ as the invariant controlling 
delocalization in the globular phase and determining unexpected
thermodynamic effects of genuinely topological nature. The outlined
scenario is totally different from that expected in the swollen
polymer regime.
Even if unusually large chain lengths need to be explored in order
to fully clarify their asymptotics, the discovered phenomena are 
most relevant in practice for finite $N$.

{\bf Acknowledgments} \\
This work is supported by
``Fondazione Cassa di Risparmio di Padova e Rovigo'' within the
2008-2009 ``Progetti di Eccellenza'' program. 



\end{document}